\documentclass[aps,pra,showpacs,superscriptaddress,twocolumn,amsmath,amssymb,longbibliography]{revtex4-1}%double-spaced,
\usepackage{amsfonts}
\usepackage{txfonts}
\usepackage{amsmath}
\usepackage{float}
\usepackage[colorlinks,breaklinks,linkcolor=blue,anchorcolor=blue,citecolor=blue,urlcolor=blue
]{hyperref}
\usepackage{graphicx}% Include figure files
\usepackage{dcolumn}% Align table columns on decimal point
\usepackage{bm}% bold math
\usepackage{amssymb}
\usepackage{mathrsfs}
\usepackage[sort&compress]{natbib}
\usepackage{subfigure}
\usepackage{braket}
\DeclareUnicodeCharacter{2009}{\,}
\begin{document}
\title{Optimizing the preparation of Dicke states using counterdiabatic driving methods}
\author{Fengzhe Tang}
\affiliation{Center for Quantum Sciences and School of Physics, Northeast Normal University, Changchun 130024, China}

\author{Gangcheng Wang}
\email{wanggc887@nenu.edu.cn}
\affiliation{Center for Quantum Sciences and School of Physics, Northeast Normal University, Changchun 130024, China}
\date{\today}

\begin{abstract}
Recently, the technique of counterdiabatic driving, which provides an effective strategy for accelerating adiabatic quantum evolution, has been widely applied in the preparation of many-body quantum states. In this work, we propose a theoretical scheme for the efficient preparation of Dicke states in a system of non-interacting two-level atoms. Our approach leverages the one-axis twisting (OAT) interaction to generate non-classical correlations and combines it with time-dependent external fields to achieve precise control over the dynamics of the system. By employing rapid adiabatic passage (RAP), it demonstrates how the system can be steered from an initial coherent spin state to a target Dicke state with high fidelity
\hypersetup{colorlinks=true,urlcolor=blue,citecolor=blue}
[S. C. Carrasco, M. H. Goerz, S. A. Malinovskaya, V. Vuleti\'c, W. P. Schleich, and V. S. Malinovsky, \href{https://doi.org/10.1103/PhysRevLett.132.153603}{Phys. Rev. Lett. \textbf{132}, 153603 (2024)}]. To further optimize the preparation process, we introduce counterdiabatic driving (CD), which suppresses non-adiabatic transitions. Numerical simulations confirm that our scheme can achieve high-fidelity Dicke states for a moderate number of particles. Our results provide a scalable and experimentally feasible approach to prepare Dicke states, with potential applications in quantum metrology, quantum communication, and quantum information processing.

\end{abstract}
\maketitle

\allowdisplaybreaks
\section{Introduction}
\label{I}
Quantum many-body systems have emerged as a powerful platform for exploring fundamental quantum phenomena and developing novel quantum technologies \cite{PhysRevB.95.094302,PhysRevLett.125.100503,PhysRevLett.111.073603,PhysRevLett.113.103004}. Among the various entangled states that can be realized in such systems, Dicke states \cite{PhysRevApplied.12.044020} hold a special place due to their unique properties and potential applications in quantum metrology, quantum communication, and quantum computing \cite{hey1999quantum,GYONGYOSI201951}. Dicke states, which are symmetric superpositions of collective excitations in an ensemble of two-level systems, exhibit strong multipartite entanglement and non-classical correlations \cite{Friedberg_2007,PhysRevLett.94.163601,PhysRevA.92.042329}. These features make them particularly valuable for high-precision measurements, quantum-enhanced sensing \cite{PhysRevA.67.012108,PhysRevA.101.012308}, and the implementation of quantum algorithms that require highly entangled states.

However, the preparation of Dicke states poses significant challenges in experimental implementations \cite{WANG2024107868,9951196,9774323,PhysRevA.104.032407,PhysRevLett.109.173604}. Traditional methods often rely on global control \cite{PhysRevA.108.012608,Brown_2023} of the system, which requires precise timing, high-fidelity operations, and sensitive measurement-based feedback \cite{PhysRevA.102.062418,PhysRevA.103.012603}. These requirements not only increase the complexity of the experimental setup but also make the system vulnerable to external noise and decoherence \cite{PhysRevA.83.013821}. Moreover, as the number of particles increases, the preparation of Dicke states becomes increasingly difficult due to the growing complexity of the system's energy structure \cite{doi:10.1021/ar000033j} and the need for precise control over the interactions between particles \cite{https://doi.org/10.1002/jcc.540110405}. These challenges have motivated researchers to explore alternative approaches that can achieve high-fidelity preparation of Dicke states while maintaining scalability and robustness against noise \cite{PhysRevA.80.052302,Campbell_2009,PhysRevA.80.052302}.

In this work, we propose a theoretical scheme that addresses these challenges by leveraging the OAT interaction \cite{PhysRevLett.132.153603,PhysRevLett.107.013601} and CD driving techniques \cite{PRXQuantum.4.010312,doi:10.1073/pnas.1619826114,PhysRevB.100.035407,Hartmann2020multispincounter,PhysRevLett.123.090602,PhysRevA.109.022201}. Our approach not only suppresses non-adiabatic transitions but also ensures robust state preparation even in the fast parameter variations. By combining RAP with CD driving, we demonstrate that our method can achieve high-fidelity Dicke states for small numbers of particles, overcoming the limitations imposed by decoherence in traditional methods. This makes our scheme particularly suitable for practical implementations in systems with strong cavity coupling, such as superconducting qubits or cold atomic ensembles \cite{doi:10.1073/pnas.1419326112}.

One promising direction is the use of collective interactions, such as the OAT interaction \cite{PhysRevLett.129.250402}, which induces spin squeezing and generates non-classical correlations essential for the preparation of Dicke states \cite{PhysRevA.106.043711}. The OAT interaction, typically realized in systems with strong nonlinearities, can create entanglement between particles without requiring direct pairwise interactions. This makes it particularly suitable for systems with a large number of particles, where direct interactions would be difficult to control. However, the OAT interaction alone is not sufficient. In recent years, RAP \cite{PhysRevA.65.043407,RANGELOV20101346,PhysRevLett.105.123003} has emerged as a powerful technique for preparing entangled states in quantum systems. RAP allows for the adiabatic transfer of population between energy levels by carefully tuning the parameters of the system \cite{PhysRevA.85.023401}, such as the frequency and amplitude of external fields. By sweeping these parameters through avoided crossings in the energy spectrum \cite{KOUSKOV1995165}, RAP can efficiently transfer the system from an initial state to a target state with high fidelity. However, the success of RAP depends critically on the adiabaticity of the process, which can be compromised by fast parameter variations or external noise. To overcome this limitation, CD driving techniques have been developed, which introduce auxiliary counterdiabatic control to suppress non-adiabatic transitions and ensure high-fidelity state preparation.

In this work, a theoretical scheme is proposed for the efficient preparation of Dicke states in a system of non-interacting two-level atoms. The approach leverages the OAT interaction to generate the necessary non-classical correlations \cite{RAHMAN2023425} and combines it with time-dependent external fields to achieve precise control over the dynamics of the system. By applying RAP, it demonstrates how the system can be steered from an initial coherent spin state to a target Dicke state. Furthermore, counterdiabatic driving is introduced to optimize the preparation process \cite{Barone_2024}, ensuring high fidelity even in the presence of non-adiabatic effects and fast parameter variations.

Our scheme is particularly well-suited for systems with strong cavity coupling \cite{Barone_2024}, where the OAT interaction can be effectively realized. Through numerical simulations, we show that our method can achieve high-fidelity preparation of Dicke states. This mechanism ensures that the prepared Dicke states remain highly coherent and robust against inhomogeneous broadening, making our scheme suitable for practical implementations in systems with fast parameter variations.

High-fidelity Dicke states can be used to enhance the precision of quantum sensors, such as atomic clocks and magnetometers, by exploiting their multipartite entanglement and non-classical correlations \cite{BELLOMO2014260,10902445}. They can also serve as a resource for quantum communication protocols, such as quantum key distribution and quantum teleportation \cite{7425090}, where the ability to generate and manipulate highly entangled states is crucial. Furthermore, our scheme provides a scalable approach to preparing Dicke states in systems, paving the way for the realization of quantum simulators and quantum computers that rely on multipartite entanglement.

The remainder of this paper is organized as follows. In Sec.~\ref{II}, we present the model and Hamiltonian, focusing on the one-axis twisting interaction and the role of time-dependent external fields. Sec.~\ref{III} introduces the counterdiabatic driving method and its application to the preparation of Dicke states. In Sec.~\ref{IV}, we analyze the preparation of Dicke states via rapid adiabatic passage and the optimization of fidelity using counterdiabatic corrections. Sec.~\ref{V} presents numerical simulations and discusses the dynamical properties of the system under different correction schemes. Finally, Sec.~\ref{VI} concludes the paper with a discussion of experimental feasibility and potential extensions.

\section{MODEL AND HAMILTONIAN}
\label{II}
\subsection{Model and Hamiltonian}
In this work, we consider a system of \(N\) non-interacting two-level atoms, where \(\hat{S}_j\) (\(j = x, y, z\)) are the components of the collective spin operators. The Hamiltonian of the system is given by:
\begin{equation}
\hat{H} = \chi \hat{S}_z^2 + \beta(t) \hat{S}_z + \Omega(t) \hat{S}_x.
\label{eq1}
\end{equation}
where \(\chi\) is the squeezing parameter associated with the OAT interaction, and \(\beta(t)\) and \(\Omega(t)\) are time-dependent external fields along the \(z\)-axis and \(x\)-axis, respectively. The OAT interaction \(\chi \hat{S}_z^2\) is capable of generating non-classical correlations and serves as a key mechanism for spin squeezing and entangled state preparation \cite{Riberi_2022}. Meanwhile, the time-dependent external fields \(\beta(t)\) and \(\Omega(t)\) provide flexible control over the system dynamics, enabling precise preparation of target states through adiabatic processes or RAP.

Here, an implementation of the RAP method for creating extreme spin squeezed states and pure Dicke states is proposed. In the weak dissipation limit \cite{PhysRevB.109.064311,PhysRevA.92.053820}, the system is well described by this Hamiltonian. This form of Hamiltonian has broad applications in practical systems, particularly in quantum information processing and quantum simulation. For instance, in superconducting qubit systems or cold atomic systems, it is widely used to prepare multipartite entangled states (e.g., Dicke states or GHZ states) and plays a crucial role in quantum metrology and quantum communication \cite{trifa:tel-04684470}.

To prepare the target state $\left| S, 0 \right\rangle$ using the RAP method, we start from the initial coherent spin state (CSS), denoted as $\left|S, S \right\rangle$. A linear frequency modulation function $\beta(t) = \alpha t u(-t)$ is applied, where the Heaviside step function $u(-t)$ ensures that the modulation rate $\beta(t)$ stops changing at $t = 0$. In this case, the linear modulation function tunes the transition between adjacent Dicke states into resonance. By appropriately choosing the driving frequency $\Omega(t)$, the system can be effectively transferred from the initial CSS state $\left| S, S \right\rangle$ to the target Dicke state $\left| S, 0 \right\rangle$. The Heaviside step function is given by:
\begin{equation}
u(t) = \begin{cases}
0 & \text{ } t < 0, \\
1 & \text{ } t \geq 0 .
\end{cases}
\end{equation}
The variation of the driving strength $\Omega(t)$ during the evolution can be divided into three stages: first, in the turn-on stage, $\Omega(t)$ gradually increases from zero to its maximum value $0.88\chi$; then, in the intermediate stage, it remains constant to maintain a stable evolution; finally, in the turn-off stage, $\Omega(t)$ gradually decreases from its maximum value to zero. This gradual design is intended to facilitate the transfer of quantum states.

The expression for $\Omega(t)$ during the turn-on and turn-off stages is:
\begin{equation}
    \Omega(t) = \frac{\Omega_{\text{max}}}{2} \left\{1 + \cos\left[\frac{\pi \left(|n t + 5| - 5\right)}{2}\right]\right\},
\end{equation}
with $\Omega_{\text{max}} = 0.88\chi$, ensuring that the drive has a consistent evolution shape during the rise and fall stages.

\subsection{Avoided Crossings and Adiabatic Passage in Energy Levels}
\begin{figure}[t]
\centering
\subfigure{
\includegraphics[width=1\columnwidth]{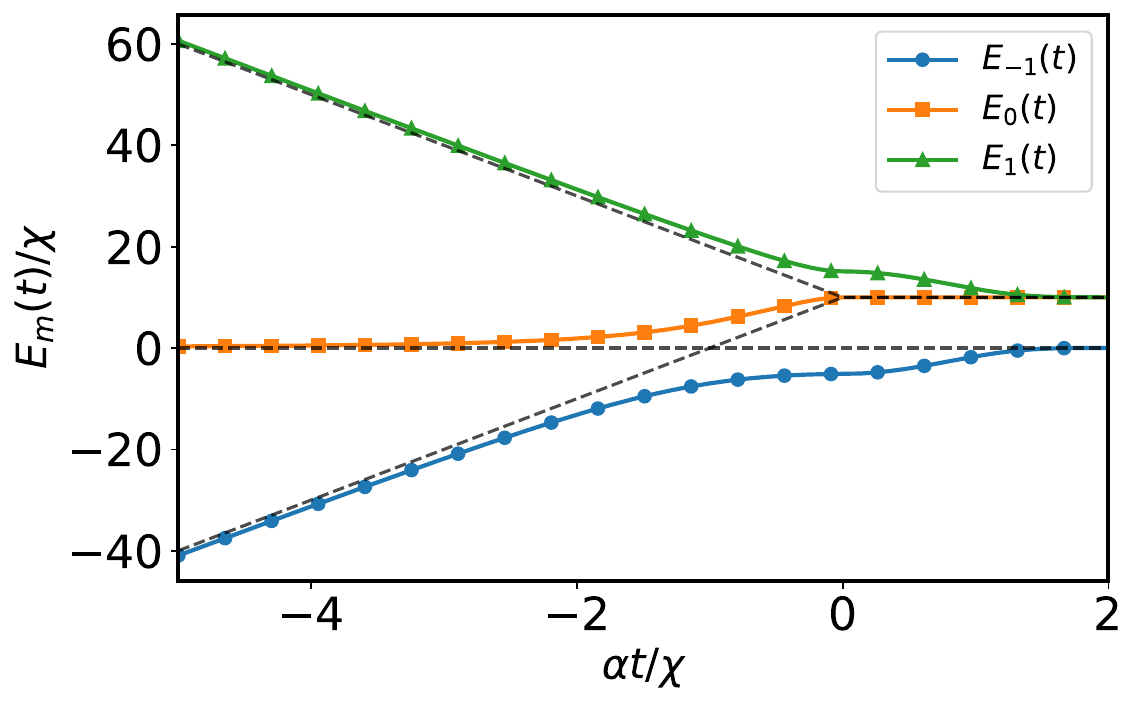}}
\caption{Five lowest Dicke state energy levels as a function of time. The solid lines represent the instantaneous eigenvalues of the Hamiltonian defined by Eq.~(\ref{eq1}), corresponding to the adiabatic picture, while the dashed lines denote the energy eigenvalues $E_m$ associated with the non-adiabatic picture. The driving field starts to decay at $t = 0$ and is completely turned off when $\alpha t/\chi = 1$. Under this condition, the chirp rate is set to $\alpha = 0.1\chi^2$, the maximum driving field is $\Omega_{\text{max}} = 0.88\chi$, and the shear parameter is $\chi = 10$.}
\label{1} % 添加标签以便引用
\end{figure}

For the situation of different energy levels, Fig.~\ref{1} shows adiabatic and non-adiabatic crossings among several energy eigenvalues $E_m(t)$. The figure shows that these crossings are transformed into avoided crossings. In Ref.~\cite{PhysRevLett.132.153603}, Carrasco et al. pointed out that the combination of the one-axis twisting Hamiltonian $\chi\hat{S}_z^2$ and the linear term $\beta(t)\hat{S}_z$ produces a unique energy level structure, forming a series of avoided-crossing channels between adjacent Dicke states that permit population transfer. The energy level diagram illustrates how the RAP method can be used to guide a multi-atom system through a sequence of avoided crossings between Dicke states. By applying the time-dependent chirped field $\beta(t)$ and the driving term $\Omega(t)$, the system population can be transferred from the initial spin coherent state $|S, S\rangle$ to the target entangled state $|S, 0\rangle$.

Fig.~\ref{1} depicts the RAP process for generating highly entangled Dicke states, where the system evolves through sequential avoided crossings between adjacent Dicke states. The dynamics are governed by a Hamiltonian comprising three key terms: $\chi \hat{S}_z^2$, $\beta(t) \hat{S}_z$, and $\Omega(t) \hat{S}_x$. Here, $\chi$ is the shearing parameter associated with the one-axis twisting interaction, which induces non-classical correlations in the system. The time-dependent chirped field $\beta(t)$ sweeps the energy levels through resonance, enabling transitions between Dicke states. Specifically, $\beta(t)$ tunes the linear Zeeman-like term $\beta(t)\hat{S}_z$, controlling the speed of the level crossings. Meanwhile, $\Omega(t)$, the driving strength associated with rotations about the $x$-axis, mediates the actual population transfer during the avoided crossings. These avoided crossings arise from the interplay between the quadratic $\hat{S}_z^2$ term and the linear $\hat{S}_z$ term, creating an effective coupling between neighboring Dicke states. Physically, the avoided crossings ensure that the system remains in the instantaneous eigenstate of the Hamiltonian, preserving coherence \cite{Wysocki:10} and preventing unwanted non-adiabatic transitions. The time scale of each crossing is determined by the chirp rate $\alpha$ and the shearing parameter $\chi$, with the period between adjacent crossings given by $\tau = 2\chi/\alpha$. By precisely controlling $\beta(t)$ and $\Omega(t)$, the system can be steered to the target Dicke state $|S,0\rangle$, which maximizes the quantum Fisher information and enables high-precision quantum metrology \cite{KOLODRUBETZ20171}.

When discussing the RAP method, by carefully designing the time dependence of $\beta(t)$ and $\Omega(t)$, the energy levels of the system can undergo avoided crossings during evolution, ensuring that the system always follows the instantaneous eigenstate and avoids non-adiabatic transitions. For example, when $\beta(t)$ varies linearly with time, the energy levels of the system gradually approach each other and undergo avoided crossings, while $\Omega(t)$ reaches its maximum value near the crossing points, driving the system from one energy level to another. This modulation scheme allows the system to adiabatically evolve from an initial state to a target state, thereby achieving high-fidelity state transfer \cite{SHEVCHENKO20101}.

\begin{figure}[t]
\centering
\subfigure{
\includegraphics[width=1\columnwidth]{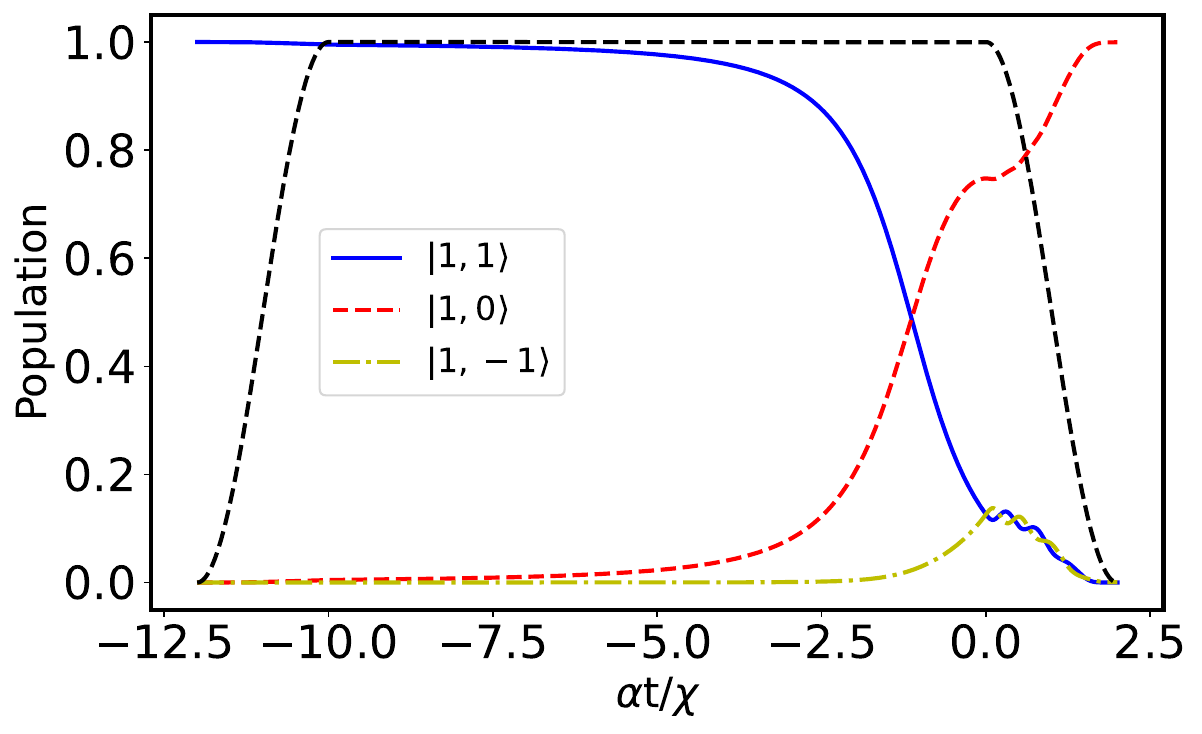}}
\caption{Population of the two-atom Dicke state as a function of the chirp rate parameter $\alpha$. The black dashed line represents the driving pulse shape $\Omega(t)$ with a chirp rate of $\alpha = 0.1\chi^2$. Other parameters are the same as Fig.~\ref{1}.}
\label{2} % 添加标签以便引用
\end{figure}

We take two atoms as an example to investigate the dynamical evolution of Dicke states. Specifically, the system is simulated with a fixed shearing parameter $\chi = 10$, and the time-dependent population dynamics of the two-atom Dicke states are plotted. In this scenario, the driving pulse is represented by the black dashed line, denoted as $\Omega(t)$. The initial chirp rate $\alpha = 0.1\chi^2$ determines the frequency modulation profile during the evolution. The final population distribution of the Dicke states is shown in Fig.~\ref{2}. When the chirp rate $\alpha = 0.1\chi^2$, the population of the target state increases significantly, while the populations of the other states decrease noticeably.

\subsection{State Preparation in the Case of Even and Odd Particle Numbers}
\begin{figure}[t]
\centering
\subfigure{
\includegraphics[width=1\columnwidth]{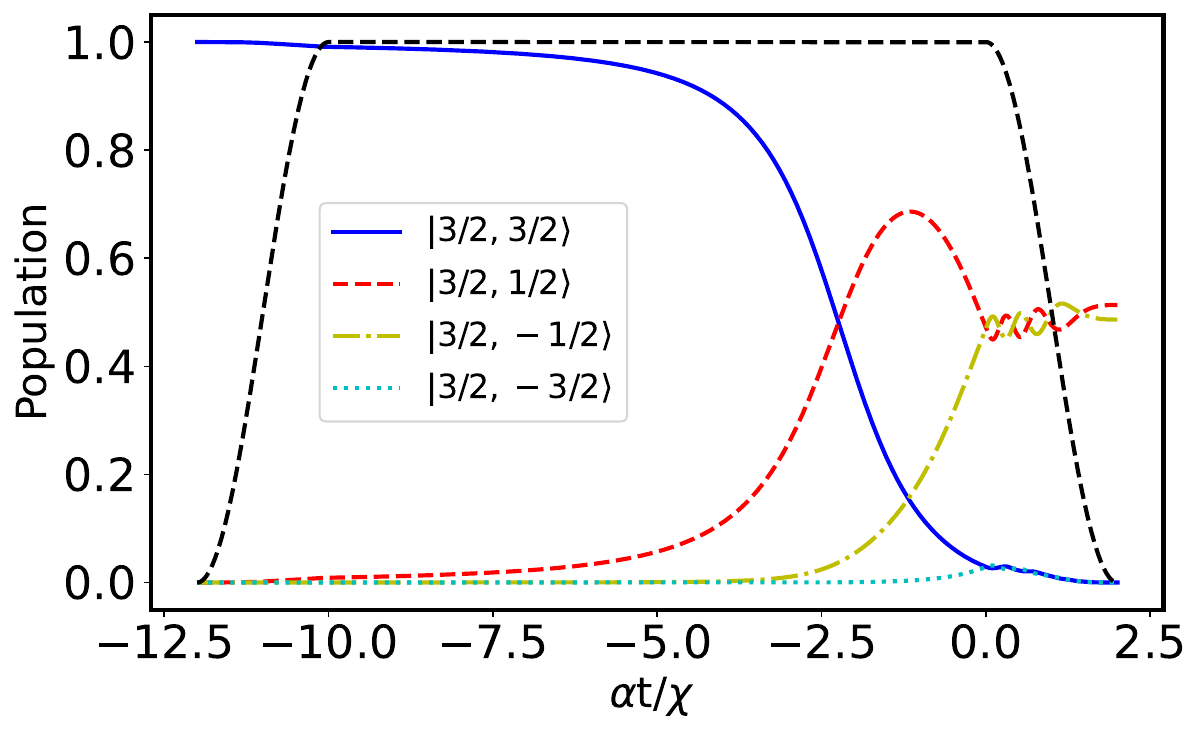}}
\caption{Population of the three-particle Dicke state as a function of the chirp rate parameter $\alpha$. The black dashed line shows the driving pulse shape $\Omega(t)$. The final state is predominantly prepared in the two lowest energy states. During the evolution, the system successfully transfers population to these two lowest energy states, while the population in other higher-energy states is significantly reduced. The chirp rate is $\alpha = 0.1\chi^2$. Other parameters are the same as Fig.~\ref{1}.}
\label{3} % 添加标签以便引用
\end{figure}

In this section, we take three atoms ($N=3$) as an example to discuss how to prepare Dicke states using an odd number of particles. In a three-atom system, the half-integer spin $S = 3/2$ yields a unique energy level structure with spin projections $m = \pm 1/2, \pm 3/2$, and there is no $\lvert S,0\rangle$ eigenstate with spin projection zero. The system evolves into a superposition of the lowest-energy states $\lvert 3/2, +1/2\rangle$ and $\lvert 3/2, -1/2\rangle$, which can become degenerate under appropriate parameter conditions \cite{Toth:07}. In the absence of the driving term $\Omega(t)  = 0$, the energy levels of these states are given by:
\begin{equation}
E_m = \chi m^2 + \beta(t) m.
\end{equation}
When $\beta(t) = 0$, the energies of the $|3/2, +1/2\rangle$ and $|3/2, -1/2\rangle$ states coincide, leading to complete degeneracy. Therefore, during the RAP process, the system populates both the $|3/2, +1/2\rangle$ and $|3/2, -1/2\rangle$ states, and the final state prepared is their symmetric superposition $(1/\sqrt{2})(\lvert 3/2, 1/2\rangle + \lvert 3/2, -1/2\rangle)$, as illustrated in Fig.~\ref{3}. According to the numerical simulation results, in the three-particle system, the population distribution among the Dicke states at the final time is approximately $0.5135$ for the $|3/2, 1/2\rangle$ state and approximately $0.4865$ for the $|3/2, -1/2\rangle$ state. The sum of the populations of the two states is close to $1$, which is in excellent agreement with the theoretical expectation (population of each basis state being $1/2$). This indicates that the system has successfully evolved into the target superposition state, verifying the effectiveness of our scheme in the three-particle system.

For odd-number particle systems, because the total spin quantum number $S$ is half-integer, the Dicke state $\lvert S,0\rangle$ does not exist. The system ultimately evolves into the symmetric superposition of the lowest-energy states $\lvert 3/2, +1/2\rangle$ and $\lvert 3/2, -1/2\rangle$. This conclusion is general: for any odd number of particles $N$, since there is no eigenstate with $m_z=0$ in the energy spectrum, the system stabilizes in the superposition of $\lvert S, \pm 1/2\rangle$ during the rapid adiabatic passage process.

\subsection{Effect of Fast Scanning}
\begin{figure}[t]
\centering
\includegraphics[width=1\columnwidth]{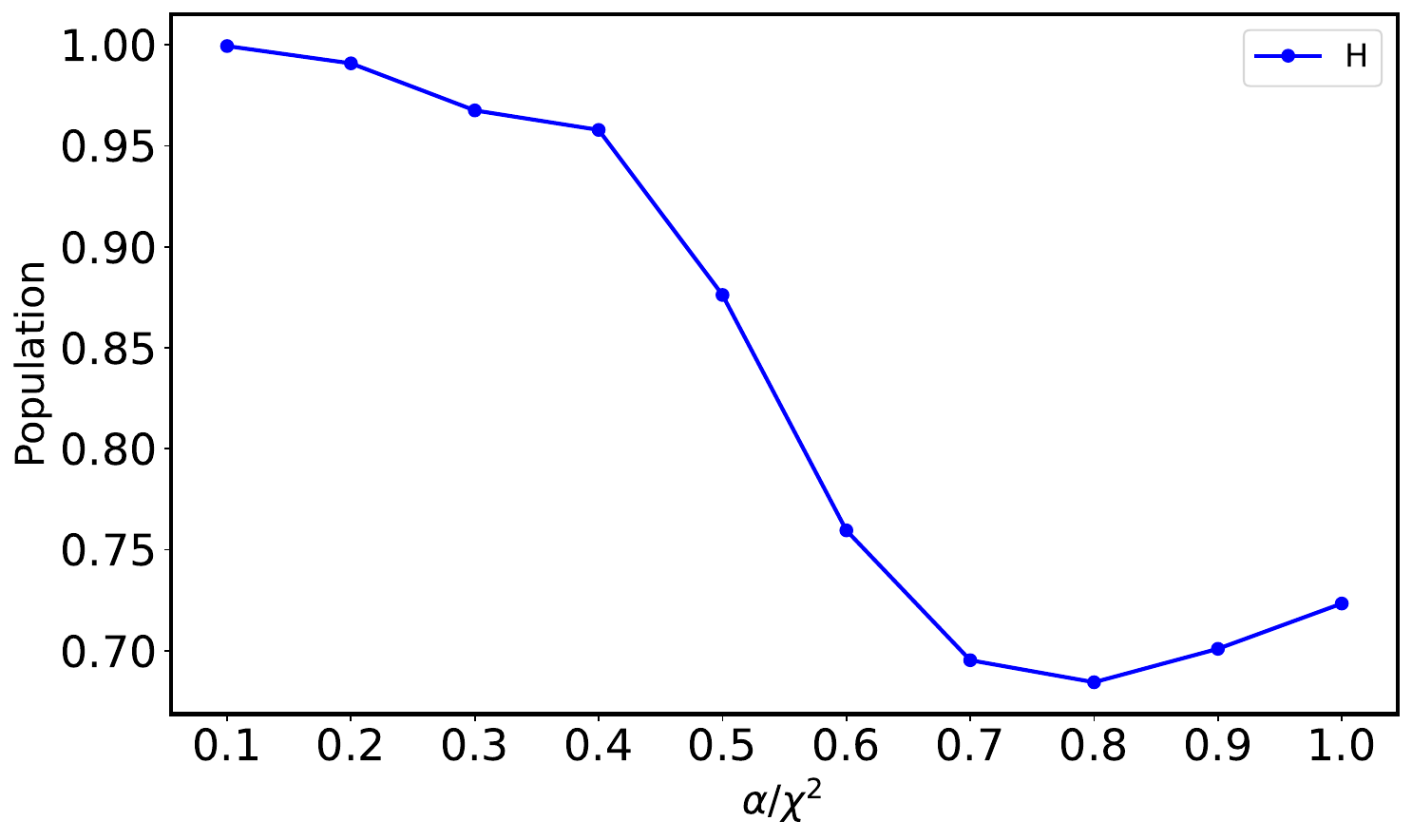}
\caption{Population of the two-atom Dicke state as a function of the chirp rate parameter $\alpha$. The blue line represents the evolution result of the original Hamiltonian. Other parameters are the same as Fig.~\ref{1}.
}
\label{4}
\end{figure}

Fig.~\ref{4} shows the result of further increasing the scanning speed in the preparation of two-atom Dicke states. When the chirp rate $\alpha$ increases, the system evolution deviates more quickly from the ideal adiabatic condition, and part of the population is transferred to non-target states, leading to a decrease in the population of the target state. This is because $\alpha$ controls the rate of change of the parameters with time. The larger $\alpha$ is, the more difficult it becomes for the system to maintain perfect adiabatic tracking \cite{doi:10.1139/A10-022,PhysRevA.90.060301}, and non-adiabatic effects are enhanced, thereby transferring population from the target state to other energy levels. Therefore, in practical applications, the chirp rate must be chosen appropriately to balance the evolution speed and the fidelity of the target state.

Increasing the chirp rate corresponds to a fast scan of the system, which may lead to a reduction in the population of the target state. To solve the problem of reduced target state population under fast scanning, a counterdiabatic driving term is introduced. That is, by adding a term to the original Hamiltonian, the system can maintain approximately adiabatic evolution even under fast scanning conditions, thereby achieving high fidelity.

\section{COUNTERDIABATIC DRIVING}
\label{III}
The core idea of counterdiabatic driving is to achieve precise control over the evolution of quantum states by modifying the system Hamiltonian, by adding a correction term to the original Hamiltonian so that the system is diagonalized in a moving reference frame \cite{PhysRevLett.123.110401}, thereby effectively suppressing non-adiabatic effects. The modified Hamiltonian is:
\begin{equation}
\hat{H}_{\mathrm{CD}}(t) = \hat{H}(\lambda) + \dot{\lambda} \hat{A}_\lambda,
\label{eqcd}
\end{equation}
The counter-diabatic driving protocol is characterized by parameter $\lambda(t)$. As long as the adiabatic gauge potential $\hat{A}_{\lambda}$ satisfies the appropriate conditions, adiabatic control can be achieved for any initial state and arbitrary driving rate (see Appendix~\ref{A} for a detailed derivation). Under the assumption of non-degenerate energy levels, its analytical solution is given by:
\begin{equation}
\langle m| \hat{A}_{\lambda}|n\rangle = i\langle m | \partial_{\lambda} n\rangle = -i \frac{\langle m| \partial_{\lambda} \hat{H}|n\rangle}{\epsilon_{m}-\epsilon_{n}},
\end{equation}
where $|n\rangle$ and $\epsilon_n$ are the eigenstates and eigenvalues of the instantaneous Hamiltonian $\hat{H}$, satisfying $\hat{H}(\lambda)|n\rangle=\epsilon_{n}|n\rangle$. The gauge potential is defined in the eigenbasis of the instantaneous Hamiltonian and requires diagonalization. Moreover, as the system size increases, $\epsilon_{m}-\epsilon_{n}$ may become exponentially small, leading to divergent matrix elements and thus making the gauge potential ill-defined in the thermodynamic limit \cite{PhysRevE.65.041303}. To address this, an approximate construction scheme for the adiabatic gauge potential \cite{PhysRevE.60.3781} is proposed, defined as follows:
\begin{equation}\label{eq6}
\hat{A}_{\lambda}^{(\ell)} = i \sum_{k=1}^{\ell} \xi_{k} \underbrace{[\hat{H},[\hat{H}, \ldots[\hat{H}}_{2 k-1}, \partial_{\lambda} \hat{H}]]].
\end{equation}
where $\xi_{k}$ are the expansion coefficients of the adiabatic gauge potential. Using a variational approximation avoids the difficulty of directly solving for the adiabatic gauge potential, and an approximate solution can be obtained by introducing variational parameters and minimizing the action. Exact solutions typically require diagonalization of the entire Hamiltonian, which is computationally prohibitive for many-body systems, making the variational approach an efficient alternative. This yields the approximate solution:
\begin{equation}
\begin{split}
\langle m | \hat{A}_{\lambda}^{(\ell)} | n \rangle &= i \sum_{k=1}^{\ell} \xi_{k}
\underbrace{\langle m | [\hat{H}, [\hat{H}, \dots [\hat{H}, \partial_{\lambda} \hat{H}] \dots ] ] | n \rangle}_{2k-1} \\
&= i \left[ \sum_{k=1}^{\ell} \xi_{k} (\epsilon_m - \epsilon_n)^{2k-1} \right]
\langle m | \partial_{\lambda} \hat{H} | n \rangle.
\end{split}
\end{equation}
Considering a finite value of $\ell$ and treating the expansion coefficients as variational parameters, these parameters can be obtained by minimizing the action $S_{\ell}$:
\begin{equation}
S_{\ell} = \operatorname{Tr}[G_{\ell}^2], \qquad G_{\ell} = \partial_{\lambda} \hat{H} - i[\hat{H}, \hat{A}_{\lambda}^{(\ell)}].
\end{equation}
Setting $\delta S_{\ell} / \delta \xi_{k} = 0$ yields the expression for $\xi_{k}$.

It should be noted that the accuracy of the variational approximation increases with the truncation order, but the computational cost also increases. In practical applications, a trade-off between accuracy and efficiency is required. Typically, low orders ($\ell = 1$ or $2$) are already sufficient to effectively suppress non-adiabatic transitions and meet the requirements for high fidelity.

\section{PREPARATION OF DICKE STATES VIA RAP AND COUNTERDIABATIC DRIVING}
\label{IV}
\subsection{Optimizing the Fidelity of Dicke State Preparation}
Through the analysis in the previous section, in the one-axis twisting model, the chirp rate affects the evolution of the system. Increasing the chirp rate corresponds to a fast sweep of the system, which may lead to a reduction in the population of the target state. To address the problem of reduced target state population under fast sweeps, a counter-diabatic driving term is introduced. That is, by adding a term to the original Hamiltonian, the system can maintain approximately adiabatic evolution even under fast sweeping conditions, thereby achieving high-fidelity state preparation.

In the intermediate driving stage, the parameter $\lambda(t)$ from the counterdiabatic driving protocol is introduced into the model, and the Hamiltonian can be written as:
\begin{equation}
\hat{H}_{a} = \chi \hat{S}_z^2 + \lambda (t) \hat{S}_z + \Omega(t) \hat{S}_x.
\end{equation}

Under the first-order approximation, when $\ell=1$, the adiabatic gauge potential can be expressed as:
\begin{equation}
\hat{A}_{\lambda}^{(1)} = i \xi_{1} [\hat{H}_{a}, \partial_{\lambda} \hat{H}_{a}] = \xi_{1}\Omega \hat{S}_{y},
\end{equation}
here $\partial_{\lambda} \hat{H}_{a} = \hat{S}_z$. $G_{1}$ is computed as:
\begin{equation}
\begin{aligned}
G_{1} &= \partial_{\lambda} \hat{H}_{a} - i[\hat{H}_{a}, \hat{A}_{\lambda}^{(1)}] \\
&= \hat{S}_z + \xi_{1}\Omega(-\chi\hat{S}_x \hat{S}_z - \chi\hat{S}_z\hat{S}_x - \lambda \hat{S}_x + \Omega\hat{S}_z).
\end{aligned}
\end{equation}
To simplify the calculation of $\operatorname{Tr}(G_1^2)$ in the Dicke subspace, we can use the following approach. Since the Hamiltonian possesses exchange symmetry (it remains invariant under the exchange of any two subsystems), we can reduce the Hamiltonian to the Dicke subspace \cite{PhysRevLett.124.063601}. This allows the computation to be carried out using the algebra of a single spin.

We can compute $G_1^2$ and take its trace in the Dicke subspace. For a two-atom system, the total Hilbert space is $2^2$-dimensional, and we only need to consider the $2j+1=3$-dimensional Dicke subspace, which simplifies the calculation. In the Dicke state subspace:
\begin{equation}
    \hat{\sigma}_D^+ = (\hat{\sigma}_D^-)^{\dagger} = \sum_{n=0}^{N-1} \sqrt{(N-n)(n+1)} \, |S,n\rangle \langle S,n+1|.
\end{equation}
Here $|S,n\rangle$ denotes the Dicke state, $N$ is the total number of particles, and $n$ is the number of particles in the spin-up state. The annihilation operator $\hat{\sigma}_D^-$ is the Hermitian conjugate of $\hat{\sigma}_D^+$. Within the Dicke subspace, the spin operators $\hat{S}_x$ and $\hat{S}_y$ are defined as $\hat{S}_x = (\hat{\sigma}_D^+ + \hat{\sigma}_D^-)/2$ and $\hat{S}_y = (\hat{\sigma}_D^+ - \hat{\sigma}_D^-)/(2i)$, while the operator $\hat{S}_z$ is a diagonal matrix whose diagonal entries correspond to the eigenvalues $n - N/2$ of $\hat{S}_z$ for the Dicke state $|S,n\rangle$. Consequently, for $N = 2$, the Hamiltonian of the system can be expressed as:
\begin{equation}
    \hat{H}_D =
    \begin{pmatrix}
        \chi + \beta & \frac{\Omega}{\sqrt{2}} & 0 \\
        \frac{\Omega}{\sqrt{2}} & 0 & \frac{\Omega}{\sqrt{2}} \\
        0 & \frac{\Omega}{\sqrt{2}} & \chi - \beta
    \end{pmatrix}.
\end{equation}
To compute $S_{1} = \operatorname{Tr}(G_{1}^{2})$ in the Dicke subspace, we need the following quantities:
\begin{equation}
\begin{aligned}
\eta &= j(j+1), \quad \mu = \eta (2j+1), \quad \operatorname{Tr}(\hat S_{L}^{2}) = \frac{1}{3} \mu, \\
\operatorname{Tr}(\hat S_{L}^{2}\hat S_{M}^{2}) &= \frac{1}{30} \mu (2\eta +1), \quad \operatorname{Tr}(\hat S_{L}\hat S_{M}\hat S_{N}) = \frac{i}{6} \mu \epsilon _{LMN}. \\[4pt]
\end{aligned}
\end{equation}
Combining the above expressions, $S_{1}$ is simplified to:
\begin{equation}
\begin{split}
S_{1} &= (1 + 2\xi_{1}\Omega^{2} + \xi_{1}^{2}\Omega^{4} + \xi_{2}^{2}\Omega^{2}\lambda^{2})\operatorname{Tr}(\hat S_{L}^{2}) \\
&\quad + 2\xi_{1}^{2}\Omega^{2}\chi^{2}\operatorname{Tr}(2\hat S_{L}^{2}\hat S_{M}^{2} + i\hat S_{L}\hat S_{M}\hat S_{N}).
\end{split}
\end{equation}

Taking the two-atom model with parameters $j = 1, \eta = 2, \mu = 6$, we obtain:
\begin{equation}
\begin{aligned}
&\operatorname{Tr}(\hat{S}_{L}^{2}) = 2, \quad \operatorname{Tr}(\hat{S}_{L}^{2} \hat{S}_{M}^{2}) = 1, \quad \operatorname{Tr}(\hat{S}_{L} \hat{S}_{M} \hat{S}_{N}) = i, \\[4pt]
&S_{1} = 2 ( 1 + 2 \xi_{1} \Omega^{2} +\xi_{1}^2 \Omega^4 + \xi_{1}^2 \Omega^2 \lambda^2 ) + 2 \xi_{1}^2 \Omega^2 \chi^2.
\end{aligned}
\end{equation}
Taking the variational derivative with respect to $\alpha_1$ and setting it to zero, i.e., $\delta S_1/\delta \xi_{1} = 0$, we obtain $\xi_{1} = -(\Omega^2 + \lambda^2 + \chi^2)^{-1}$. Thus, the first-order adiabatic gauge potential is $\hat{A}_{\lambda}^{(1)} = -\Omega \hat{S}_y(\Omega^2 + \lambda^2 + \chi^2)^{-1}$. Substituting this expression into the first-order counterdiabatic Hamiltonian results in:
\begin{equation}
    \hat{H}_{\rm CD}^{(1)} = \hat{H}_{a} + \dot{\lambda} \hat{A}_{\lambda}^{(1)} = \hat{H}_{a} - \dot{\lambda} \frac{\Omega \hat{S}_y}{\Omega^2 + \lambda^2 + \chi^2}.
\end{equation}

Similarly, the second-order approximation for the intermediate driving stage (see Appendix~\ref{B}) yields the following values for $\xi_{1}$ and $\xi_{2}$:
\begin{equation}
\begin{aligned}
\xi_{1} &= -\bigl(8 \lambda^4 + 2 \Omega^4 + 8 \chi^2 \lambda^2 + \chi^2 \Omega^2 + 37 \lambda^2 \Omega^2\bigr) \\
&\qquad  \bigl(4 \lambda^6 + \Omega^6 - 8 \chi^2 \lambda^4 + 4 \chi^4 \lambda^2 + 33 \lambda^2 \Omega^4 \\
&\qquad + 36 \lambda^4 \Omega^2 + 28 \chi^2 \lambda^2 \Omega^2\bigr)^{-1}, \\[4pt]
\xi_{2} &= \bigl(4 \lambda^2 + \Omega^2\bigr)  \bigl(4 \lambda^6 + \Omega^6 - 8 \chi^2 \lambda^4 + 4 \chi^4 \lambda^2 \\
&\quad + 33 \lambda^2 \Omega^4 + 36 \lambda^4 \Omega^2 + 28 \chi^2 \lambda^2 \Omega^2\bigr)^{-1}.
\end{aligned}
\end{equation}

To facilitate the calculation of the counterdiabatic correction terms for the turn-on, intermediate, and turn-off stages, we denote the nested commutator part in Eq.~(\ref{eq6}) as $\hat{O}_{2k-1}$, satisfying:
\begin{equation}
    \left\{
    \begin{aligned}
        \hat{O}_k &= [\hat{H}_{a}, \hat{O}_{k-1}], \quad k \geq 1 \\
        \hat{O}_0 &= \partial_{\lambda} \hat{H}_{a}.
    \end{aligned}
    \right.
\end{equation}
Thus, Eq.~(\ref{eq6}) can be expanded as:
\begin{equation}
    \hat{A}_{\lambda}^{(\ell)} = i\hbar \sum_{k=1}^{\ell} \xi_{k} \hat{O}_{2k-1}.
\end{equation}
$G_{\lambda}^{(\ell)}$ can be expressed as:
\begin{equation}
    G_{\lambda}^{(\ell)} = \partial_{\lambda} \hat{H}_{a} + \frac{i}{\hbar} [ \hat{A}_{\lambda}^{(\ell)}, \hat{H}_{a} ] = \hat{O}_0 + \sum_{k=1}^{\ell} \xi_{k} \hat{O}_{2k},
\end{equation}
and the cost function $S_\ell(\vec{\xi})$ takes the form:
\begin{equation}
    S_\ell(\vec{\xi_{k}}) = \operatorname{Tr}[\hat{O}_0^2] + 2 \sum_{k=1}^{\ell} \xi_{k} \operatorname{Tr}[\hat{O}_0 \hat{O}_{2k}] + \sum_{j,k=1}^{\ell} \xi_{j} \xi_{k} \operatorname{Tr}[\hat{O}_{2j} \hat{O}_{2k}].
\end{equation}
We can define $\vec{\xi} = (\xi_{k}, \ldots, \xi_\ell)$ and express $S_\ell$ as a quadratic polynomial:
\begin{equation}
    S_\ell(\vec{\xi}) = A + 2 \vec{B} \cdot \vec{\xi} + \vec{\xi}^T C \cdot \vec{\xi}.
\label{eq2}
\end{equation}
The specific forms of the coefficients $A$, $\vec{B}$, and the matrix $C$ are given by:
\begin{equation}
    A = \operatorname{Tr}[\hat{O}_0^2],
\end{equation}
\begin{equation}
    B_i = \operatorname{Tr}[\hat{O}_0 \hat{O}_{2i}],
\end{equation}
\begin{equation}
    C_{ij} = \operatorname{Tr}[\hat{O}_{2i} \hat{O}_{2j}].
\end{equation}
We introduce a real matrix $U$ that diagonalizes $C$, satisfying $D = U^T C U$. Then $S_\ell$ can be rewritten as:
\begin{equation}
    S_\ell(\vec{\xi}') = A + \sum_{k=1}^{\ell} (2 B'_k \xi'_k + D_{kk} {\xi'_k}^2),
\end{equation}
where $\vec{\xi}' = U^T \vec{\xi}$, $\vec{B}' = U^T \vec{B}$. Setting $\delta S_\ell(\vec{\xi}')/\delta \xi_k' = 0$ yields $\xi_k' = -B_k' D_{kk}^{-1}$. For $\ell = 1$, the minimum of the quadratic action $S_1(\vec{\xi})$ is given by Eq.~\eqref{eq2}:
\begin{equation}
    \xi_1 = -\frac{\operatorname{Tr}\left[\hat{O}_0 \hat{O}_2\right]}{\operatorname{Tr}\left[\hat{O}_2^2\right]}.
\end{equation}

Using this method, we substitute the operators $\hat{O}_0$ and $\hat{O}_2$ for the turn-on, intermediate, and turn-off stages to obtain the corresponding correction terms. If the chirp rate $\alpha$ increases, then $n = \alpha / (0.1\chi^2)$. Here $n$ represents the scanning multiplication factor, which increases the scanning speed and correspondingly shortens the scanning time. The evolution time runs from $-12\chi/\alpha$ to $2\chi/\alpha$.

In the turn-on stage, we set $\lambda = \Omega(t)$, and thus only consider corrections related to $\Omega(t)$ while ignoring those related to $\beta(t)$. This is because $\Omega(t)$ plays the dominant role in driving population transfer during the turn-on stage, while the contribution of $\beta(t)$ is relatively small. Under our chosen parameters, this simplifies the analysis while maintaining the accuracy of state preparation.
In the turn-on stage, the time-dependent parameters and the first-order adiabatic gauge potential are given by:
\begin{equation}
\begin{aligned}
\lambda &= \frac{\Omega_{\text{max}}}{2} \left\{1 + \cos\left[\frac{\pi (-n t - 10)}{2}\right]\right\}, \\
\xi_1 &= -\frac{\beta^2 + \chi^2}{\beta^4 + 6 \beta^2 \chi^2 + \beta^2 \lambda^2 + \chi^4 + 4 \chi^2 \lambda^2}, \\
\hat{A}_{\lambda}^{(1)} &= -\xi_{1} \left[ \chi (\hat{S}_z \hat{S}_y + \hat{S}_y \hat{S}_z) + \beta \hat{S}_y \right].
\end{aligned}
\end{equation}
Here $\lambda$ describes the time-dependent modulation of the external field in the turn-on stage, and $\xi_1$ is the first-order variational parameter obtained by minimizing the action $S_1$. $\hat{A}_{\lambda}^{(1)}$ is the first-order adiabatic gauge potential, which suppresses non-adiabatic transitions during the turn-on stage.

In the intermediate stage, $\Omega(t)$ remains constant. In the turn-off stage, the step function results in $\beta = 0$. Then we set $\lambda = \Omega(t)$, analogous to the turn-on stage. The final Hamiltonian including counterdiabatic driving can be computed via Eq.~(\ref{eqcd}).

\section{Numerical Simulation and Dynamical Properties}
\label{V}
To further improve the fidelity of the prepared state based on the RAP method, the counterdiabatic driving approach is employed \cite{Do_2022}, and the counterdiabatic term is introduced to optimize the evolution process of the system. After adding the correction term to the system Hamiltonian, a high population in the target state can be maintained even when the chirp rate increases. This is because the counterdiabatic driving compensates for the deviation from the ideal adiabatic condition, enabling the system to maintain a high population in the target state. This approach allows for fast evolution while mitigating the effects of decoherence, thereby improving the fidelity of the target state.

\begin{figure}[t]
\centering
\includegraphics[width=1\columnwidth]{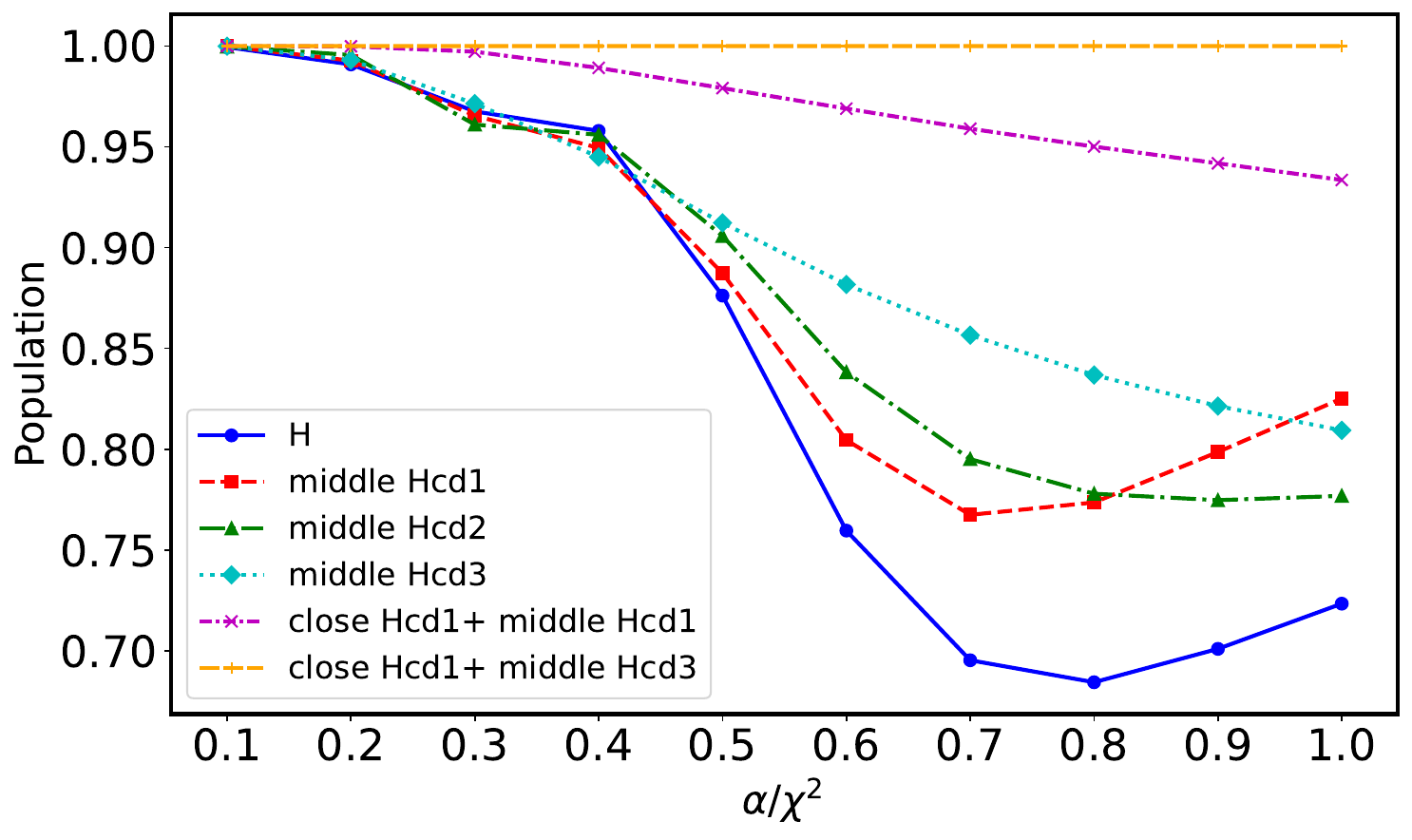}
\caption{Population of the two-atom Dicke state as a function of the chirp rate $\alpha$. The blue line (H) is the uncorrected reference; the red line (middle Hcd1) represents the first-order correction in the intermediate stage of the drive; the dark green line (middle Hcd2) and light green line (middle Hcd3) represent the second-order and third-order corrections in the intermediate stage, respectively; the purple line (close Hcd1 + middle Hcd1) combines the first-order corrections of both the closing and intermediate stages; the orange line (close Hcd1 + middle Hcd3) combines the first-order correction of the closing stage with the third-order correction of the intermediate stage. Other parameters are the same as Fig.~\ref{1}.}
\label{5}
\end{figure}

Fig.~\ref{5} shows the evolution of the final population of the target Dicke state as a function of the chirp rate $\alpha$ under different counterdiabatic correction terms. The figure is used to clarify the influence of different corrections on the population. It can be seen that, compared with the uncorrected reference case, introducing counterdiabatic corrections in the intermediate or turn-off stages leads to significant changes in the population. Moreover, combinations of corrections of different orders can further increase the population of the target state, indicating that both the stage at which the correction is applied and the order of the correction play a role in regulating the population transfer efficiency. Furthermore, as can be seen from Fig.~\ref{5}, for the two-atom system, after introducing the first-order counterdiabatic correction in the turn-off stage and the third-order correction in the intermediate stage, the population of the target Dicke state remains stable above $0.99$. This result fully verifies the effectiveness of the combined correction scheme \cite{6093757} in suppressing nonadiabatic transitions, and demonstrates that high-fidelity state preparation can still be achieved under fast evolution conditions by introducing counterdiabatic driving of different orders in different stages.

Further analysis of the two-atom system shows that the improvement achieved by adding the counterdiabatic driving term in the turn-on stage is very limited. The key reason is that during the turn-on stage, the driving strength $\Omega(t)$ gradually increases from zero to its maximum value, and the system population is mainly concentrated in the initial coherent spin state, with no large-scale interlevel transfer having yet occurred. The nonadiabatic effect \cite{doi:https://doi.org/10.1002/0471433462.ch7} itself is relatively weak, and the counterdiabatic correction lacks a clear object to act upon, making it difficult to produce a significant effect. In stark contrast, during the turn-off stage, $\Omega(t)$ rapidly decays from its peak value, and the system is in the critical stage where the population converges to the target Dicke state. The rapid parameter change induced by the chirp rate $\alpha$ significantly enhances the probability of nonadiabatic transitions, leading to population loss from the target state to intermediate or higher energy states. The counterdiabatic correction term precisely compensates for the nonadiabatic deviation in the evolution of the energy levels, effectively suppressing this population loss and thus greatly increasing the population in the target state. It is worth noting that although extending the duration of the turn-off stage can also suppress nonadiabatic effects by reducing the rate of parameter change, this would directly prolong the preparation cycle, reducing the efficiency of state preparation and potentially introducing more decoherence noise due to the increased interaction time with the environment, thereby affecting the final fidelity. Therefore, introducing the counterdiabatic driving term in the turn-off stage can effectively suppress the loss of target state population while maintaining a relatively fast evolution speed, thereby balancing efficiency and fidelity. Its improvement effect is significantly better than applying the correction only in the intermediate stage, where $\Omega(t)$ remains constant and the system has essentially entered the adiabatic evolution regime, with the nonadiabatic effect being weaker than in the turn-off stage, leaving relatively limited room for compensation by the correction term.

\begin{figure}[t]
\centering
\includegraphics[width=1\columnwidth]{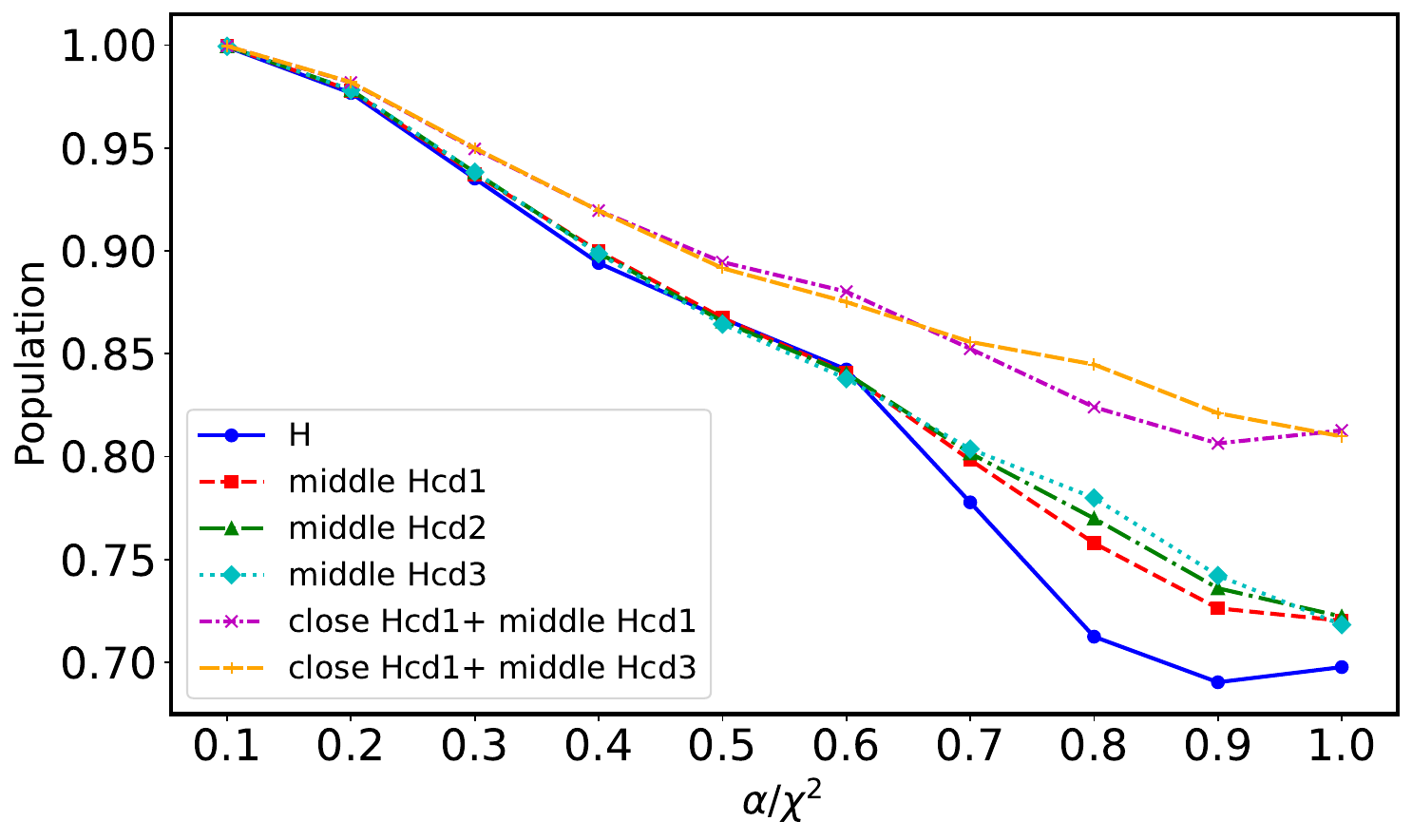}
\caption{Population of the four-particle Dicke state as a function of the chirp rate $\alpha$. Different lines represent various scenarios where counter-diabatic terms are added at different stages of the drive. Other parameters are the same as Fig.~\ref{1}.}
\label{6}
\end{figure}

Extending the system to four particles, Fig.~\ref{6} further verifies the above regularities: in both the two-atom and four-atom systems, adding counterdiabatic terms in the turn-off stage significantly increase the population transfer efficiency, while corrections in the turn-on stage have negligible effects and are not shown in the figure. It is noteworthy that as the number of particles increases, the effect of counterdiabatic correction in the intermediate stage becomes less pronounced. The physical reason is that with increasing particle number, the density of energy levels of the Dicke states increases significantly, and the spacing between adjacent energy levels gradually narrows. In the intermediate stage, where $\Omega(t)$ remains constant, the system evolution has already approached a stable adiabatic trajectory. In this regime, the main source of nonadiabatic transitions is no longer the rapid variation of parameters but rather small perturbations caused by energy level degeneracy or near-degeneracy. Since the counterdiabatic correction term is designed to compensate for nonadiabatic deviations induced by rapid parameter changes, its ability to suppress such perturbations is limited. Furthermore, as the number of particles increases, the Hilbert space dimension of the system grows exponentially, and the population distribution in the intermediate stage becomes more dispersed. A single correction term is insufficient to cover all nonadiabatic transition channels, which further weakens the effectiveness of the correction in the intermediate stage. In contrast, the core issue in the turn-off stage always remains the population loss caused by rapid parameter changes. This physical mechanism does not change with increasing particle number, so the counterdiabatic correction in the turn-off stage consistently maintains a significant improvement effect.

\section{DISCUSSION AND CONCLUSION}
\label{VI}
In this work, taking the Hamiltonian with one-axis twisting interaction as the object, the first- and second-order approximations of the adiabatic gauge potential are derived, and corresponding control strategies are designed for the turn-on, intermediate, and turn-off stages of system evolution. The avoided crossing property of the system is verified through energy level diagrams, confirming that the proposed scheme can guide the system to achieve stable population transfer and exhibits good scalability.

Focusing on the preparation of Dicke states, this chapter uses the Hamiltonian with one-axis twisting interaction as the core model and proposes a preparation scheme based on RAP combined with counterdiabatic driving. The system consists of \(N\) non-interacting two-level atoms as the research object. Through the joint modulation of time-dependent fields \(\beta(t)\) and \(\Omega(t)\), the transfer from the initial CSS state to the target Dicke state \(\left| S, 0 \right\rangle\) is achieved. The results show that the RAP method can achieve high population in the target state at low chirp rates, while under fast scanning conditions, non-adiabatic effects lead to a significant decrease in population. To address this, counterdiabatic driving is introduced for correction. Experimental feasibility analysis shows that introducing counterdiabatic correction in the turn-off stage has the most significant effect on improving the population of the target state, followed by the intermediate stage, while the effect in the turn-on stage is limited. This regularity holds for both two-atom and four-atom systems, verifying the applicability of the proposed scheme in these systems, and providing an experimentally feasible technical path for achieving fast evolution and high-fidelity preparation of Dicke states.

\section*{Acknowledgments}
This work is supported by the Innovation Program for Quantum Science and Technology (No. 2023ZD0300700).

\appendix
\renewcommand{\appendixname}{APPENDIX}
\section{DERIVATION OF ADIABATIC GAUGE POTENTIAL}
\label{A}

$\hat{H}$ is the Hamiltonian in the laboratory frame, and the Hamiltonian in the moving frame takes the form:
\begin{equation}
 \hat {\widetilde {H}} = \hat U^\dagger \hat H \hat U + i \dot{\hat{U}}^\dagger\hat U,
\label{a1}
\end{equation}
where $\hat{U}$ is the unitary operator that transforms from the laboratory frame to the moving frame, which depends on the parameter $\lambda$ ($\lambda$ is a function of time), satisfying $\dot{\hat{U}} = \partial_\lambda \hat{U} \dot{\lambda}$. Since $\hat{U}$ satisfies the unitarity condition $\hat{U}^\dagger \hat{U}=\hat{I}$, we have:
\begin{align}
\frac{d}{dt}(\hat{U}^\dagger \hat{U}) &= \dot{\hat{U}}^\dagger \hat{U} + \hat{U}^\dagger \dot{\hat{U}} = 0.
\end{align}
From the above equation we obtain $\dot{\hat{U}}^\dagger = -\dot\lambda \hat U^\dagger \partial_\lambda \hat U$. The adiabatic gauge potential in the moving frame is defined as $\hat{\widetilde{A}}_\lambda = i \hat{U}^\dagger \partial _\lambda \hat U$.
The adiabatic gauge potential in the laboratory frame is $\hat{A}_\lambda = i  \partial _\lambda$.
To diagonalize the Hamiltonian in the moving frame, the second term in Eq.~(\ref{a1}) must be eliminated. For this purpose, we need to add a term to the Hamiltonian in the laboratory frame, namely:
\begin{equation}
\hat{H}_{\mathrm{CD}}(t) = \hat{H}(\lambda) + \dot{\lambda} \hat{A}_\lambda.
\end{equation}

Next, we discuss the analytical solution of the adiabatic gauge potential. First, in the eigenbasis of $\hat{H}(\lambda)$, the diagonal elements of the adiabatic gauge potential are actually the Berry connection:
\begin{equation}
A_\lambda^{(n)} = \langle n(\lambda) | \hat{A}_\lambda | n(\lambda) \rangle = i \langle n(\lambda) | \partial_\lambda | n(\lambda) \rangle.
\end{equation}
For different eigenstates $|m\rangle$ and $|n\rangle$ (with $m \neq n$), there exists the identity $\langle m | \hat{H}(\lambda) | n \rangle = 0$.
Under the adiabatic approximation, the Hamiltonian $\hat{H}(\lambda)$ is diagonalized among its eigenstates. This means that when the system is in a particular eigenstate, it does not spontaneously transition to another eigenstate unless disturbed by an external perturbation. Therefore, for two different eigenstates $|m\rangle$ and $|n\rangle$, the transition amplitude (the matrix element of the Hamiltonian) is zero.

Next, we differentiate this identity with respect to the parameter $\lambda$ to obtain:
\begin{equation}
\begin{aligned}
0 &= \langle \partial_\lambda m | \hat{H} | n \rangle + \langle m | \partial_\lambda \hat{H} | n \rangle + \langle m | \hat{H} | \partial_\lambda n \rangle \\
&= \epsilon_n \langle \partial_\lambda m | n \rangle + \epsilon_m \langle m | \partial_\lambda n \rangle + \langle m | \partial_\lambda \hat{H} | n \rangle \\
&= (\epsilon_m - \epsilon_n) \langle m | \partial_\lambda n \rangle + \langle m | \partial_\lambda \hat{H} | n \rangle.
\end{aligned}
\end{equation}
Through the above derivation, we obtain:
\begin{equation}
\langle m| \hat{A}_{\lambda}|n\rangle = i\langle m | \partial_{\lambda} n\rangle = -i \frac{\langle m| \partial_{\lambda} \hat{H}|n\rangle}{\epsilon_{m}-\epsilon_{n}}.
\end{equation}
This equation shows the relationship between the adiabatic gauge potential and the derivative of the Hamiltonian with respect to the parameter for different eigenstates.

\section{SECOND-ORDER COMMUTATORS WITH OPERATORS NOTATION}
\label{B}
Next, we calculate the second-order adiabatic gauge potential expression. For $\ell = 2$, adiabatic gauge potential $\hat{A}^{(2)}_\lambda = i \xi_1 [\hat{H}_a, \partial_\lambda \hat{H}_a] + i \xi_2 [\hat{H}_a, [\hat{H}_a, [\hat{H}_a, \partial_\lambda \hat{H}_a]]]$.  the following commutators are required:
\begin{equation}
\begin{aligned}
[\hat{H}_a, \partial_\lambda \hat{H}_a] &= -i \Omega \hat{S}_y, \\[4pt]
[\hat{H}_a, [\hat{H}_a, \partial_\lambda \hat{H}_a]] &= -\Omega \chi \hat{S}_z \hat{S}_x - \Omega \chi \hat{S}_x \hat{S}_z
- \Omega \lambda \hat{S}_x + \Omega^2 \hat{S}_z, \\[4pt]
[\hat{H}_a, [\hat{H}_a, [\hat{H}_a, \partial_\lambda \hat{H}_a]]] &= -\Omega \chi^2 \hat{S}_z^2 \hat{S}_y - \Omega \chi^2 \hat{S}_z \hat{S}_y \hat{S}_z \\
&\quad- (1+i) \Omega \chi \lambda (\hat{S}_z \hat{S}_y - \hat{S}_y \hat{S}_z) - i \Omega \lambda^2 \hat{S}_y.
\end{aligned}
\end{equation}
From this we obtain:
\begin{equation}
\begin{aligned}
G_2 &= \partial_\lambda \hat{H}_a + i [ \hat{A}_\lambda^{(2)}, \hat{H}_a ] \\
&= - ( \xi_1 \Omega \chi + 3 \xi_2 \Omega \chi \lambda^2
+ 2 \xi_2 \Omega^3 \chi ) \hat{S}_z \hat{S}_x - ( \xi_1 \Omega \chi + 3 \xi_2 \Omega \chi \lambda^2 \\
&\quad+ 2 \xi_2 \Omega^3 \chi ) \hat{S}_x \hat{S}_z
 - (\xi_1 \Omega \lambda + \xi_2 \lambda^3 \Omega + \xi_2 \Omega^3 \chi ) \hat{S}_x + ( 1 + \xi_1 \Omega^2\\
&\quad  + \xi_2 \Omega^2 \lambda^2 + \xi_2 \Omega^4 ) \hat{S}_z
- \xi_2 \Omega \chi^3 \hat{S}_z^3 \hat{S}_x- 3\xi_2 \Omega \chi^3 \hat{S}_z^2 \hat{S}_x \hat{S}_z \\
&\quad - 3 \xi_2 \Omega \chi^2 \lambda \hat{S}_z^2 \hat{S}_x + 4 \xi_2 \Omega^2 \chi^2 \hat{S}_z^3 - 5\xi_2 \Omega^2 \chi^2 \hat{S}_z \hat{S}_y^2 \\
&\quad- 2\xi_2 \Omega^2 \chi^2 \hat{S}_y \hat{S}_z \hat{S}_y - 3\xi_2 \Omega \chi^3 \hat{S}_z \hat{S}_x \hat{S}_z^2
- \xi_2 \Omega \chi^3 \hat{S}_x \hat{S}_z^3 \\
&\quad- 3\xi_2 \Omega \chi^2 \lambda \hat{S}_x \hat{S}_z^2 - 5 \xi_2 \Omega^2 \chi^2 \hat{S}_y^2 \hat{S}_z - 6\xi_2 \Omega \chi^2 \lambda \hat{S}_z \hat{S}_x \hat{S}_z \\
&\quad- 6\xi_2 \Omega^2 \chi \lambda \hat{S}_y^2  + 4 \xi_2 \Omega^2 \chi \lambda \hat{S}_z^2 + 2 \xi_2 \Omega^2 \chi^2 \hat{S}_z \hat{S}_x^2 \\
&\quad+ 2 \xi_2 \Omega^2 \chi^2 \hat{S}_x^2 \hat{S}_z + 2 \xi_2 \Omega^2 \chi \lambda \hat{S}_x^2.
\end{aligned}
\end{equation}
Next, we compute $S_2$, and then determine the optimal values of $\xi_1$ and $\xi_2$ by setting the variation to zero, $S_2 = \operatorname{Tr}(G_2^2)$. To find the optimal values of $\xi_1$ and $\xi_2$, we solve the following system of equations: $\delta S_2 / \delta \xi_1 = 0, \quad \delta S_2 / \delta \xi_2 = 0.$ The solution yields:
\begin{equation}
\begin{aligned}
\xi_{1} &= -\bigl(8 \lambda^4 + 2 \Omega^4 + 8 \chi^2 \lambda^2 + \chi^2 \Omega^2 + 37 \lambda^2 \Omega^2\bigr) \\
&\qquad  \bigl(4 \lambda^6 + \Omega^6 - 8 \chi^2 \lambda^4 + 4 \chi^4 \lambda^2 + 33 \lambda^2 \Omega^4 \\
&\qquad + 36 \lambda^4 \Omega^2 + 28 \chi^2 \lambda^2 \Omega^2\bigr)^{-1}, \\[4pt]
\xi_{2} &= \bigl(4 \lambda^2 + \Omega^2\bigr)  \bigl(4 \lambda^6 + \Omega^6 - 8 \chi^2 \lambda^4 + 4 \chi^4 \lambda^2 \\
&\quad + 33 \lambda^2 \Omega^4 + 36 \lambda^4 \Omega^2 + 28 \chi^2 \lambda^2 \Omega^2\bigr)^{-1}.
\end{aligned}
\end{equation}
Substituting the above results into $\hat{A}^{(2)}_\lambda$, we obtain its explicit expression:
\begin{equation}
\begin{split}
\hat{A}^{(2)}_\lambda = & (\Omega \xi_1 + \xi_2 \Omega \lambda^2 + \xi_2 \Omega^3) \hat{S}_y
+ \xi_2 \Omega \chi^2 (\hat{S}_z^2\hat{S}_y + \hat{S}_y\hat{S}_z^2 + 2\hat{S}_z\hat{S}_y\hat{S}_z) \\
&+ 2\xi_2 \Omega \chi \lambda (\hat{S}_z\hat{S}_y + \hat{S}_y\hat{S}_z) - \xi_2 \Omega^2 \chi (\hat{S}_x\hat{S}_y + \hat{S}_y\hat{S}_x).
\end{split}
\end{equation}
The second-order counterdiabatic Hamiltonian $\hat{H}_{\mathrm{CD}}^{(2)}$ is defined as:
\begin{equation}
\hat{H}_{\mathrm{CD}}^{(2)} = \chi \hat{S}_z^2 + \lambda \hat{S}_z + \Omega \hat{S}_x + \dot{\lambda} \hat{A}^{(2)}_\lambda.
\end{equation}

\bibliography{ref}
\providecommand{\noopsort}[1]{}\providecommand{\singleletter}[1]{#1}%
\end{document}